\documentclass[aps,prb,10pt,twocolumn,showpacs,amsmath,amssymb,superscriptaddress,floatfix,longbibliography]{revtex4-2}
\usepackage[T1]{fontenc}
\usepackage{graphicx}
\usepackage{color}
\usepackage[colorlinks,bookmarks=false,citecolor=blue,linkcolor=red,urlcolor=blue]{hyperref}
\usepackage{multirow}
\usepackage{ulem}
\usepackage{tabularx}
\usepackage[nounderscore]{syntax}
\usepackage{comment}

\graphicspath{{fig/}}
\usepackage{subfigure}
\usepackage[percent]{overpic}

\newcommand{\be}{\begin{equation}}
\newcommand{\ee}{\end{equation}}

\begin{document}

\title{Eigenoperator Entanglement Statistics in Local Lindbladians}

\author{Ze-Kai Hong}
\affiliation{School of Physics, Peking University, Beijing 100871, China}

\author{Xu Feng}
\affiliation{School of Physics, Peking University, Beijing 100871, China}

\author{Zhi-Cheng Yang}
\email{ zcyang19@pku.edu.cn}
\affiliation{School of Physics, Peking University, Beijing 100871, China}
\affiliation{Center for High Energy Physics, Peking University, Beijing 100871, China}

\date{\today}
\begin{abstract}
Random matrix spectral statistics are widely used to diagnose quantum chaos in open systems, but whether chaotic eigenvalue correlations are accompanied by Haar-typical eigenoperators remains unclear. We study the operator-entanglement statistics of Liouvillian eigenoperators with near-maximal entanglement, analogous to states near the middle of the spectrum in Hamiltonian systems. Using the Kullback–Leibler divergence, we show that Haar-random statistics accurately describe both the Ginibre and class-$\mathrm{AI}^{\dagger}$ Gaussian ensembles. For purely dissipative random Lindbladians, we find that locality of the jump operators drastically alters the eigenoperator-entanglement statistics. Nonlocal Lindbladians remain relatively close to Haar-random behavior, whereas local Lindbladians deviate increasingly from random-matrix statistics with system size. Moreover, unlike in Hamiltonian systems, the most highly entangled eigenoperators of local Lindbladians do not generally lie in the region of largest density of states, owing to the clustered structure of the complex eigenspectrum. We introduce an iterative $\sigma$-clipping scheme to extract the high-entanglement distribution without preselecting a spectral window, and find that it is Gaussian for nonlocal Lindbladians but log-normal for local ones. Remarkably, the \textit{same} log-normal distribution, with no additional fitting, also describes dissipative mixed-field Ising chains belonging to distinct non-Hermitian symmetry classes. Our results therefore point to a universal eigenoperator-entanglement distribution for local Lindbladians that is not captured by generic non-Hermitian random-matrix ensembles.
\end{abstract}
\maketitle

\section{Introduction}
\label{intro}

Random matrix theory (RMT) provides one of the central frameworks for diagnosing
quantum chaos. In closed quantum systems, chaotic Hamiltonians exhibit
universal level correlations described by the Wigner--Dyson distribution~\cite{BGS,Atas2013}, 
while their highly excited eigenstates typically satisfy the
eigenstate thermalization hypothesis (ETH) and exhibit volume-law
entanglement with random-state-like statistical properties
~\cite{Deutsch1991,Srednicki1994,DAlessio2016,VidmarRigol2017,EntanglementVedika}.

For open quantum systems governed by Lindblad master equations, the
corresponding spectral problem is generically non-Hermitian, and the
Liouvillian eigenvalues are distributed in the complex plane. The use of
complex spectral correlations as a diagnostic of dissipative quantum chaos
dates back to the pioneering works of Grobe, Haake, and Sommers, who
demonstrated a quantum distinction between regular and chaotic dissipative
motion~\cite{GHS} and subsequently established universal cubic level
repulsion in dissipative quantum chaos~\cite{GHS2}. Later studies extended this picture to many-body Lindbladians and showed that spectral correlations in chaotic regimes are often well described by non-Hermitian random-matrix ensembles, including the Ginibre ensembles~\cite{Akemann2019,SciPostPhysCore.5.2.026,PhysRevA.105.L050201,Hamazaki2020,CSR}.
In particular, the complex spacing ratio provides an
unfolding-free probe that distinguishes random-matrix-like correlations
from uncorrelated complex spectra~\cite{CSR}. Complementary studies of
random Lindblad generators have further characterized their global
spectral distributions, spectral gaps, and steady-state properties
~\cite{Denisov2019,Can2019,SaRandom2020,Costa2023}, contributing to a broader
understanding of dissipative quantum chaos and its modern diagnostic
framework~\cite{PhysRevLett.127.170602,PhysRevLett.130.140403,SaRibeiroDenisov2026}.

Nonetheless, physical Lindbladians possess additional structures that are not captured by generic non-Hermitian random-matrix ensembles. The situation is analogous to, but potentially richer than, that of physical Hamiltonians, where locality introduces additional constraints that can lead to drastic differences between the bulk and edges of the spectrum. Indeed, recent studies have revealed much richer structures in the complex spectra of physical Lindbladians than those of generic random matrices. For example, purely dissipative random Lindbladians, among the least structured models of open quantum systems, exhibit a universal lemon-shaped spectral profile in the complex plane, in contrast to the circular spectrum of Ginibre random matrices~\cite{Denisov2019}. Imposing locality on the jump operators introduces further structure, leading to spectral clustering in the complex plane organized by the operator weights of the corresponding eigenoperators~\cite{Wang2020,Hartmann2024, ChirameBurnell2026}.

On the other hand, the eigensystem of a Hamiltonian or Liouvillian contains not only its eigenvalues, but also its eigenstates (or eigenoperators, in the case of Liouvillians). Over the past few years, a growing body of work has explored the statistical properties of eigenstates as a potentially finer characterization of quantum chaos, including the statistics of the von Neumann entanglement entropy~\cite{VidmarRigol2017,Bianchi2022,EntanglementVedika} and the full entanglement spectrum~\cite{PhysRevLett.115.267206,PhysRevB.96.020408,PhysRevB.93.174202}. More recently, Ref.~\cite{EntanglementVedika} introduced a metric that quantifies the distance between the microcanonical distribution of eigenstate entanglement entropies and that of a Haar-random ensemble, providing a more refined probe of the degree of quantum chaos. A central finding is that the entanglement distribution of mid-spectrum eigenstates of local Hamiltonians is well described by random matrix theory once energy conservation is properly taken into account. For local Lindbladians, it was shown that the bulk eigenoperators exhibit a constrained form of randomness: their total weights depend strongly on Pauli-string size and on the associated decay rate, while the coefficients within a fixed operator-size sector are nearly maximally scrambled~\cite{ChirameBurnell2026}. However, it remains unclear whether, and to what extent, the entanglement properties of Liouvillian eigenoperators are captured by generic non-Hermitian random matrix ensembles.

In this work, we study the distribution of operator entanglement entropy among Liouvillian eigenoperators with near-maximal entanglement, analogous to eigenstates near the middle of a Hamiltonian spectrum, where the entanglement is typically largest. Using the Kullback–Leibler (KL) divergence introduced in Ref.~\cite{EntanglementVedika}, we quantify how closely the highly entangled eigenoperators of a given Liouvillian resemble Haar-random vectors. As a benchmark, we first show that, for the three random-matrix ensembles relevant to this work—the real and complex Ginibre ensembles and the class-$\mathrm{AI}^{\dagger}$ ensemble (complex matrices satisfying $M=M^T$)~\cite{Hamazaki2020}—the eigenvector entanglement entropy is well described by a Gaussian distribution whose mean and variance agree with those of Haar-random states. 

We next turn to purely dissipative random Lindbladians with either nonlocal or local jump operators. Here, an immediate subtlety arises that has no counterpart in Hamiltonian or Floquet systems. The spectrum of a Lindbladian occupies the two-dimensional complex plane, so there is no natural analog of the mid-spectrum eigenstates associated with a real spectrum. Moreover, eigenoperators with the highest entanglement entropy generally do not lie in regions with the largest density of states in the complex plane. It is therefore difficult to define an analog of a microcanonical energy window and extract the mean and variance of the entanglement entropy in a model-independent and unbiased manner.
To address this difficulty, we introduce an iterative selection scheme inspired by the $\sigma$-clipping method in statistics to extract the high entanglement distribution {\it without preselecting a spectral window}. At each step, we discard eigenoperators whose entanglement entropy is sufficiently low relative to the maximal value and repeat the procedure until the retained set converges. We then evaluate a two-moment approximation to $D_{\rm KL}$, determined by the mean and variance of the resulting distribution. For nonlocal Lindbladians, the entanglement-entropy distribution remains relatively close to that of the Haar-random ensemble, with the residual discrepancy arising primarily from deviation in the mean entropy. For local Lindbladians, by contrast, the resulting $D_{\rm KL}$ increases rapidly with system size, indicating a pronounced departure from Haar-random behavior. More strikingly, while the converged distribution is well approximated by a Gaussian for nonlocal Lindbladians, it instead follows a log-normal distribution for local Lindbladians.
We further test these findings in a physical mixed-field Ising chain with two classes of local dissipators, which place the resulting Lindbladians in the complex Ginibre (GinUE) and ${\rm AI}^\dagger$ symmetry classes, respectively. Remarkably, we find that the normalized entanglement entropy distribution follows the {\it same} log-normal distribution as in the random Lindbladian models, {\it without} any refitting. Our results therefore provide strong evidence for a universal eigenoperator-entanglement distribution in local Lindbladians.

On the other hand, the level-spacing statistics of all models considered in this work show excellent agreement with random-matrix predictions for the corresponding symmetry classes. Our results thus reveal potentially universal distinctions between physical Lindbladians and generic non-Hermitian random matrices that are invisible to level-spacing probes.

\section{Eigenvalue and eigenoperator statistics of Liouvillians}

\subsection{Spectral statistics}\label{sec:csr}

The evolution of open quantum systems in the Markovian regime is usually described by a quantum master equation $\partial_t \rho = \mathcal{L}(\rho)$, where $\mathcal{L}$ is the Liouvillian superoperator, also called the Liouvillian or Lindbladian. It is convenient to represent operators $A$ acting on a $D$-dimensional Hilbert space as a vector $|A\rangle\rangle$ in a $D^2$-dimensional Hilbert space, such that the Hilbert-Schmidt inner product of two operators is given by the inner product of the corresponding vectors:
\begin{equation}
{\rm Tr}(A^\dagger B) = \langle\langle A|B\rangle\rangle.
\end{equation}
Upon vectorizing the density matrix, the quantum master equation takes a form analogous to the Schr\"odinger equation in a doubled Hilbert space $\partial_t |\rho\rangle\rangle = \mathcal{L}|\rho\rangle\rangle$, where the Liouvillian is cast as a $D^2\times D^2$ non-Hermitian matrix. The dynamical properties of the system are thus completely encoded in the eigenvalues and eigenvectors of the Liouvillian. For the eigenvalues, a useful probe of universal correlations is provided by the level-spacing statistics. For Hermitian matrices with real eigenvalues, one considers the distribution of spacings between adjacent levels, $s_i=\epsilon_{i+1}-\epsilon_i>0$. Uncorrelated levels obey Poisson statistics, whereas chaotic spectra exhibit Wigner--Dyson statistics with power-law level repulsion, $P(s)\sim s^{\beta}$ as $s\rightarrow 0$, where the exponent $\beta$ is determined by the symmetry class. Level-spacing statistics have therefore become a standard diagnostic for distinguishing quantum-chaotic systems from integrable or localized ones.

For open quantum systems, the eigenvalues of the Liouvillian are generally complex. One can similarly define the level spacing as the Euclidean distance between nearest-neighboring eigenvalues in the complex plane. A more robust probe of spectral correlations that does not require unfolding is provided by the complex spacing ratio,
\begin{align}
z_{j}=\frac{\lambda_{j}^{\mathrm{N N}}-\lambda_{j}}{\lambda_{j}^{\mathrm{N N N}}-\lambda_{j}},
\end{align}
where $\lambda_{j}^{\mathrm{N N}}$ and $\lambda_{j}^{\mathrm{N N N}}$ are the nearest and next-nearest neighbors of the eigenvalue $\lambda_{j}$ in the complex plane. Since the spacing ratio is itself complex, $z=re^{i\phi}$, spectral correlations are characterized by both its radial and angular distributions~\cite{CSR}. In particular, the first moments $(\langle r\rangle,-\langle\cos\phi\rangle)$ take the values $(2/3,0)$ for uncorrelated spectra, $(0.739,0.247)$ for the Ginibre ensembles, and $(0.722,0.195)$ for the class-$\mathrm{AI}^\dagger$ ensemble~\cite{CSR}. Below, we use the complex spacing ratio as a spectral diagnostic of quantum chaos for all models considered in this work.

\subsection{von Neumann entanglement entropy statistics}\label{sec:ee-statistics}

Spectral statistics capture only half of the RMT phenomenology; the other half concerns eigenvectors, which correspond to eigenoperators in open quantum systems. In recent years, von Neumann entanglement-entropy statistics have been used to compare the ensemble properties of eigenstates in physical models with those of RMT ensembles~\cite{Bianchi2022,EntanglementVedika}.
For each right eigenoperator $R_{\alpha}$, normalized by
$\operatorname{Tr}(R_{\alpha}^{\dagger}R_{\alpha})=1$, we expand
\begin{align*}
 |R_{\alpha}\rangle\!\rangle
 =
 \sum_{\mathbf{m}}c_{\mathbf{m}}^{(\alpha)}
 |\mathbf{m}\rangle\!\rangle
\end{align*}
in the normalized Pauli-string basis. For a spatial bipartition
$N=N_A+N_B$, with $N_A=\lfloor N/2\rfloor$, the coefficients are reshaped
into a matrix
$c^{(\alpha)}\in\mathbb{C}^{4^{N_A}\times 4^{N_B}}$. We then define
\begin{align*}
 \rho_A^{(\alpha)}
 &=
c^{(\alpha)}c^{(\alpha)\dagger},&
 S_A^{(\alpha)}
 &=
 -\operatorname{Tr}
 \left[
 \rho_A^{(\alpha)}\ln\rho_A^{(\alpha)}
 \right].
\end{align*}
Throughout this work, $S_A$ denotes this half-chain spatial operator
entanglement entropy of the right eigenoperators. It should not be confused
with an entanglement bipartition between the original and copied Hilbert
spaces introduced by vectorization.
We quantify deviations from RMT by computing the KL divergence,
\begin{align}
 D_{\mathrm{KL}}(P_{\mathrm{E}},P_{\mathrm{R}})=\int dS_{A}P_{\mathrm{E}}(S_{A})\log\frac{P_{\mathrm{E}}(S_{A})}{P_{\mathrm{R}}(S_{A})}\geq\, 0,
\end{align}
where $P_{\mathrm{E/R}}(S_{A})$ denotes the entanglement-entropy distribution of the model and the reference ensemble, respectively. We model the reference Page distribution obtained from a Haar-random ensemble as Gaussian, with the finite-size Page mean $\mu_{\mathrm{R}}$~\cite{Page1993} and variance $\sigma_{\mathrm{R}}^2$. This is a good approximation, as higher moments for the Haar random ensemble are further suppressed by $1/D$~\cite{PhysRevE.93.052106, PhysRevE.96.022106, PhysRevD.100.105010}.

We make no Gaussian assumption about $P_\mathrm{E}(S_A)$. Instead, we replace it by a moment-matched Gaussian with the same mean $\mu_{\mathrm{E}}$ and variance $\sigma_{\mathrm{E}}^2$. Under this approximation, the integral reduces to the following expression:
\begin{align}
D_{\mathrm{KL}}\approx  \underbrace{\frac{(\mu_{\mathrm{E}}-\mu_{\mathrm{R}})^{2}}{2\sigma_{\mathrm{R}}^{2}}}_{D_\mu}+ \underbrace{\frac{1}{2}\left[  \frac{\sigma_{\mathrm{E}}^{2}}{\sigma_{\mathrm{R}}^{2}} -1-\log\frac{\sigma_{\mathrm{E}}^{2}}{\sigma_{\mathrm{R}}^{2}} \right]}_{D_\sigma},\label{DKL-gaussian}
\end{align}
which gives a moment-based lower bound on the KL divergence.
We denote the first and second terms in Eq.~\eqref{DKL-gaussian} by
$D_{\mu}$ and $D_{\sigma}$, respectively, and refer to them as the mean
contribution and the variance contribution. The mean contribution measures the displacement of the model mean from the finite-size Page mean, whereas the variance contribution measures the mismatch between the model and Page variances.

\section{Models}
We now summarize the models considered in this work. We study two classes of Liouvillians: purely dissipative models with random jump operators and  physical spin chains with local dissipators.

\subsection{Purely dissipative models with random Lindbladians}\label{sec:random-lindbladian}

A generic Liouvillian in the absence of a coherent Hamiltonian evolution can be written in the following Gorini-Kossakowski-Sudarshan-Lindblad form~\cite{Lindblad1,Lindblad2,BreuerPetruccione,Lidar2019}
\begin{align}\label{eq:RandomLindblad}
 \mathcal{L}_{D}[\rho]= & \sum_{\mu, \nu=1 }^{M(w_{\mathrm{max}})}R_{\mu \nu}\left( L_{\mu}\rho L^{\dagger}_{\nu}-\frac{1}{2} \{ L_{\nu}^{\dagger}L_{\mu},\rho \}\right),
\end{align}
where ${L_\mu}$'s form a traceless Hermitian operator basis, satisfying
$L_\mu^\dagger=L_\mu$ and normalized according to
$\operatorname{Tr}(L_\mu^\dagger L_\nu)=\delta_{\mu\nu}$. The Kossakowski matrix $R$ is a complex positive semidefinite random matrix. We choose the operators $L_{\mu}$ to be normalized Pauli strings on a system of $N$ sites. A Pauli string $\hat{P}$ is the normalized direct product $\frac{1}{\sqrt{ 2^{N} }}\bigotimes_{j=1}^{N}\sigma_{j}^{m_{j}}$, where $m_{j}\in\{ 0,1,2,3 \}$, $\sigma_{j}^{0}=\mathbb{I}_{2\times 2}$, and $\sigma_{j}^{1,2,3}=\sigma_{j}^{x,y,z}$ are the standard $2\times 2$ Pauli matrices acting on site $j$. The weight $w(\hat{P})$ of a Pauli string counts the number of non-identity operators in the string and is defined as
\begin{align}
w(\hat{P})=\mathrm{Tr}\left[\hat{P}\mathcal{S}[\hat{P}]\right],
\end{align}
where the map $\mathcal{S}$ is defined as
\begin{align}
 \mathcal{S}[\hat{P}]=\frac{1}{4}\sum_{i=1}^{N}\sum_{m_{i}=1}^{3}(\hat{P}-\sigma_{i}^{m_{i}}\hat{P}\sigma_{i}^{m_{i}}).
\end{align}
In constructing the model, we take the jump operators to be the set of all Pauli strings with total weight $w\leq w_{\max}$. To further explore the effect of locality, we consider two cases: (i) nonlocal Lindbladians with $w_{\max}=N$; and (ii) local Lindbladians with $w_{\max}=2$. Here ``local'' refers to locality in Pauli-string complexity, rather than spatial locality: the non-identity support of an included Pauli string need not occupy contiguous sites. For a given $w_{\max}$, the number of included traceless Pauli strings, and hence the dimension of the jump-operator space, is
\begin{equation}
 M(w_{\max})
 =
 \sum_{t=1}^{w_{\max}}
 3^{t}\binom{N}{t}.
\end{equation}
The Kossakowski matrix $R$ is therefore an $M(w_{\max})\times M(w_{\max})$ matrix. We generate it as $R=U\mathsf{D}_{R}U^{\dagger}$, where $\mathsf{D}_{R}$ is diagonal with entries sampled independently from a fixed positive uniform distribution and $U$ is a random unitary matrix of dimension $M(w_{\max})$ drawn from the circular unitary
ensemble.  We then rescale each realization to satisfy $\operatorname{Tr}R=2^{N}$. This convention removes the arbitrary overall scale of the sampled diagonal entries, leaving $w_{\max}$ as the only adjustable model parameter.

\subsection{Mixed-field Ising chain with local dissipators}
\label{sec:ising}

The second model we consider contains both coherent evolution and dissipation. The corresponding Liouvillian takes the form
\begin{equation}\label{eq:LME}
\mathcal{L}(\rho) = -i[H,\rho] + \sum_a L_a \rho L_a^\dagger - \frac{1}{2} \big\{ L_a^\dagger L_a, \rho \big\},
\end{equation}
where $L_a$'s denote the jump operators. Compared with Eq.~(\ref{eq:RandomLindblad}), 
Eq.~(\ref{eq:LME}) is written in the diagonal Lindblad representation obtained by
diagonalizing the Kossakowski matrix, with its eigenvalues absorbed into
the jump operators. We take the Hamiltonian to be a one-dimensional mixed-field Ising model (MFIM) with open boundary conditions.
\begin{align}
 H=-J\sum_{j=1}^{N-1}\sigma_{j}^{z}\sigma_{j+1}^{z}-h_{x}\sum_{j=1}^{N}\sigma_{j}^{x}-h_{z}\sum_{j=1}^{N} \sigma_{j}^{z}.
\end{align}
For the dissipative part, we consider two classes of local jump operators, chosen such that the resulting Liouvillians possess different symmetries and belong to distinct non-Hermitian random-matrix universality classes~\cite{Hamazaki2020,SaSymmetry2023,Kawabata2023}. In the first class, the local jump operators are taken to be~\cite{ChirameBurnell2026}:
\begin{equation}
    \begin{aligned}
 L^{(1)}_{j}= & \sqrt{ \gamma }\sigma_{j}^{+}\quad &&j=1,\dots,N \\
 L^{(2)}_{j}= & \frac{\sqrt{ \gamma }}{2}\sigma^z_{j} \quad &&j=1,\dots N\\
 L^{(3)}_{j}= & \frac{\sqrt{ \gamma }}{4}(\Bbb{I}+\sigma^x_{j})(\Bbb{I}+\sigma^x_{j+1}) \quad &&j=1,\dots N-1 \\
 L^{(4)}_{j}= & \frac{\sqrt{ \gamma }}{4}(\Bbb{I}+\sigma^x_{j})(\Bbb{I}+\sigma^x_{j+2}) \quad &&j=1,\dots N-2.
\end{aligned}
\label{eq:ginue}
\end{equation}
These four sets of channels describe local incoherent pumping, on-site
dephasing, and correlated short-range dissipation acting on nearest- and
next-nearest-neighbor pairs, respectively. The resulting Liouvillian does not have any additional symmetry, and hence belongs to the GinUE symmetry class.
In the second class, the dissipators contain only single-site and two-site dephasing:
\begin{equation}
\begin{aligned}
  L_{j}^{(1)}= & \sqrt{ \gamma^{(1)}_{j} }\sigma^z_{j}\quad &&j=1,\dots,N \\
 L_{j}^{(2)}= & \sqrt{ \gamma^{(2)} }\sigma^z_{j}\sigma^z_{j+1}\quad &&j=1,\dots,N-1,
\end{aligned}
\label{eq:AI}
\end{equation}
where $\gamma_{j}^{(1)}$ is taken to be a random variable. In the conventional computational basis, the MFIM Hamiltonian satisfies $H^{T}=H$, and every jump operator in Eq.~(\ref{eq:AI}) satisfies $L_{\mu}^{T}=L_{\mu}$ and $[L_{\mu}^{\dagger},L_{\mu}]=0$. The vectorized Liouvillian therefore obeys the transposition symmetry $\mathcal{L}^{T}=\mathcal{L}$, corresponding to $\mathrm{TRS}^{\dagger}$ with square $+1$ and hence symmetry class $\mathrm{AI}^{\dagger}$~\cite{Hamazaki2020}. These models test the robustness of our findings for random Liouvillians in a more physically realistic setting.

For numerical simulations, we set $J=1$, $h_x=1.3$, and $h_z=1.2$ for the MFIM Hamiltonian. For the GinUE-MFIM, all four dissipative channels in Eq.~(\ref{eq:ginue}) have the same rate, $\gamma=0.8$. For the ${\rm AI}^\dagger$-MFIM, the single-site dephasing rates $\gamma_j^{(1)}$ are drawn independently from a uniform distribution on the interval $[0,0.8]$, while the two-site dephasing rate is fixed at $\gamma^{(2)}=0.8$.

\section{Random Lindbladians}\label{sec:Pure-Dissipations}

This section establishes the central numerical observation of this work: random Lindbladians can display Ginibre-like eigenvalue statistics without exhibiting the corresponding Haar-like eigenoperator statistics. We first calibrate the entanglement diagnostic on GinUE and class-$\mathrm{AI}^{\dagger}$ Gaussian random matrices. We then apply the spectral and eigenoperator entanglement diagnostics to the purely dissipative model introduced in Sec.~\ref{sec:random-lindbladian}. We will see that locality in the Lindbladian can drastically alter the distribution of operator entanglement entropy in the near-maximal-entanglement regime. Moreover, the absence of a well-defined analog of a microcanonical energy window for complex spectra motivates the iterative $\sigma$-clipping scheme introduced below, which allows us to select highly entangled eigenoperators without imposing an arbitrary spectral window.

\subsection{Eigenoperator entanglement distribution for random matrix ensembles}

The real and complex Ginibre ensembles, denoted by GinOE and GinUE, consist of
matrices with independent real and complex Gaussian entries, respectively. The
class-$\mathrm{AI}^{\dagger}$ Gaussian ensemble instead consists of
complex-symmetric matrices satisfying $M^{T}=M$. For GinUE, bi-unitary
invariance gives each normalized right eigenvector a complex-Haar marginal on
the unit sphere. This result is reviewed in Ref.~\cite{ByunForrester2025} and underlies broader eigenvector-delocalization
results~\cite{RudelsonVershynin2016,LytovaTikhomirov2020}. Therefore, the eigenoperator
entanglement of GinUE random matrices follows the finite-size complex-Haar Page distribution. By contrast, GinOE is invariant under orthogonal similarity,
$P_{\mathrm{GinOE}}(M)=P_{\mathrm{GinOE}}(OMO^{T})$ for
$O\in O(4^{N})$, whereas the Gaussian class-$\mathrm{AI}^{\dagger}$ ensemble
is invariant under unitary congruence,
$P_{\mathrm{AI}^{\dagger}}(M)=P_{\mathrm{AI}^{\dagger}}(UMU^{T})$ for
$U\in U(4^{N})$~\cite{ByunForrester2025,AkemannAygunKieburgPaessler2025}.
These symmetries do not establish a complex-Haar marginal for generic right
eigenvectors, and we are not aware of analytical results on the distribution of the eigenstate entanglement in these ensembles.
Therefore, we numerically compute the $D_{\rm KL}$ between the entanglement distributions of GinOE as well as class-${\rm AI}^\dagger$ ensembles and the Page distribution.

\begin{figure}[!t]
    \centering
    \begin{overpic}[width=\columnwidth]{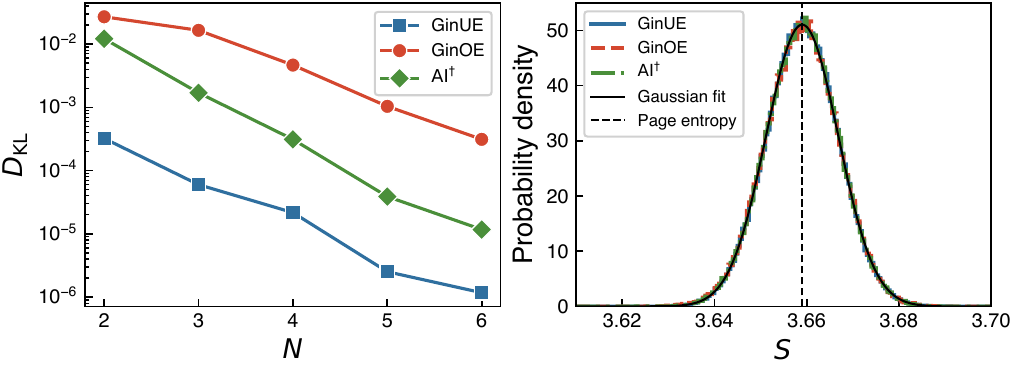}
            \put(14,8.5){\small (a)}
            \put(60,8.5){\small (b)}
        \end{overpic}
    \caption{(a) The $D_{\rm KL}$ of Eq.~(\ref{DKL-gaussian}) between the half-system entanglement entropy distributions of GinUE, GinOE, and Gaussian class-$\mathrm{AI}^{\dagger}$ right eigenvectors and the finite-size complex-Haar Page distribution as a function of system size $N$. For $N=2,\ldots,6$, each ensemble contains $500$, $300$, $100$, $150$, and $50$ realizations, respectively. (b) Full entanglement distributions of the three ensembles at $N=6$. The solid black curve shows a Gaussian fit to the GinUE distribution only, while the vertical black dashed line marks the finite-size Page entropy.}
    \label{fig:ginibre-ee-benchmark}
\end{figure}

For each realization of random matrices, we use all $4^N$ right eigenvectors of the matrix, and compare their half-chain entanglement entropy distributions with the Page reference using the Gaussian-approximation KL divergence Eq.~(\ref{DKL-gaussian}). Fig.~\ref{fig:ginibre-ee-benchmark}(a) shows the resulting $D_{\mathrm{KL}}$ as a function of $N$. We find that $D_{\rm KL}$ decreases exponentially with system size for all three random matrix ensembles. In particular, GinUE is already close to the complex-Haar reference at the smallest sizes and reaches values of order $10^{-6}$ at the largest system size. GinOE exhibits larger finite-size corrections, but $D_{\rm KL}$ still decreases from order $10^{-2}$ to order $10^{-4}$. The Gaussian class-$\mathrm{AI}^{\dagger}$ ensemble likewise approaches the complex-Haar reference, with $D_{\mathrm{KL}}$ decreasing from order $10^{-2}$ to order $10^{-5}$. In Fig.~\ref{fig:ginibre-ee-benchmark}(b), we further show the full entanglement-entropy distributions for all three random-matrix ensembles. The three distributions are nearly indistinguishable and are well described by a Gaussian whose mean is given by the Page entropy. These results indicate that the eigenstate-entanglement distribution is largely insensitive to the underlying symmetry class, and that the Page distribution therefore provides a meaningful benchmark for all models considered in this work.

\subsection{Spectral statistics of the purely dissipative random Lindbladians}

\begin{figure}[t]
    \centering
        \begin{overpic}[width=0.98\columnwidth]{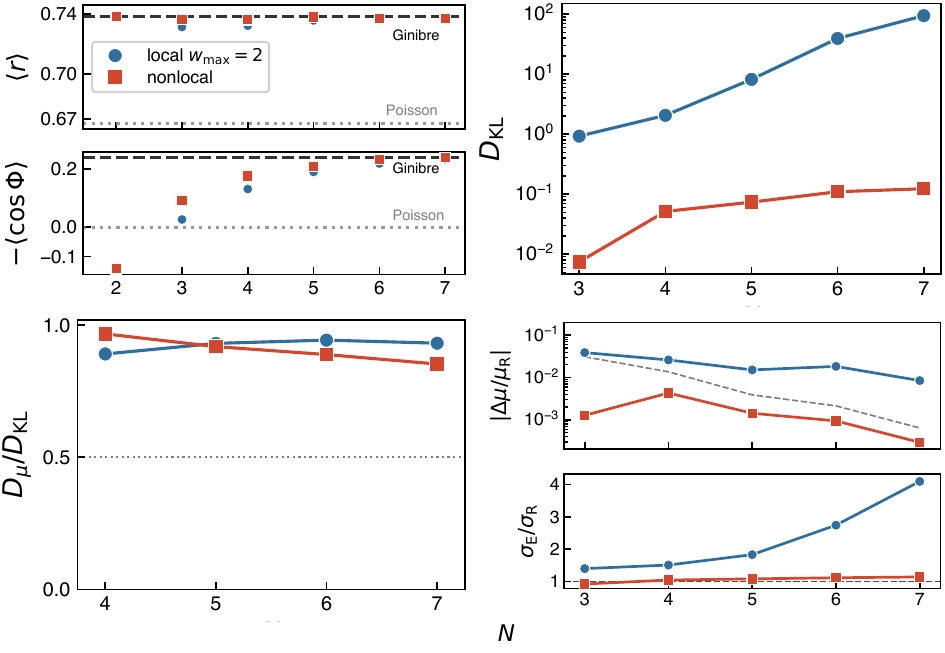}
            \put(30,59){\small (a)}
            \put(30,42){\small (b)}
            \put(62,60){\small (c)}
            \put(10,10){\small (d)}
            \put(64,22.5){\small (e)}
            \put(64,13){\small (f)}
        \end{overpic}
    \caption{Spectral and eigenoperator-entanglement statistics for the purely dissipative random Lindbladians. (a)\&(b) Complex spacing ratio for random Lindbladians with either nonlocal ($w_{\rm max}=N$, red squares) or local $(w_{\rm max}=2$, blue circles) jump operators. (a) Average $\langle r\rangle$; (b) average $-\langle {\rm cos}\phi\rangle$. Horizontal dashed lines indicate the GinUE and Poisson reference values. (c) Gaussian-approximation KL divergence~(\ref{DKL-gaussian}) between the eigenoperator-entanglement distribution and the Page reference distribution. (d) Fraction of $D_{\mathrm{KL}}$ arising from the mean contribution, $D_{\mu}/D_{\mathrm{KL}}$. (e) Absolute relative deviation of the mean EE from the Page mean, $|\Delta\mu/\mu_{\mathrm{R}}|$, where $\Delta\mu\equiv\mu_{\mathrm{E}}-\mu_{\mathrm{R}}$. The gray dashed curve shows $\sigma_{\mathrm{R}}/\mu_{\mathrm{R}}$, so values above this curve correspond to $|\Delta\mu|>\sigma_{\mathrm{R}}$. (f) Ratio of the standard deviation to the Page standard deviation, $\sigma_{\mathrm{E}}/\sigma_{\mathrm{R}}$.  }
    \label{fig:model-a-csr-dkl}
\end{figure}
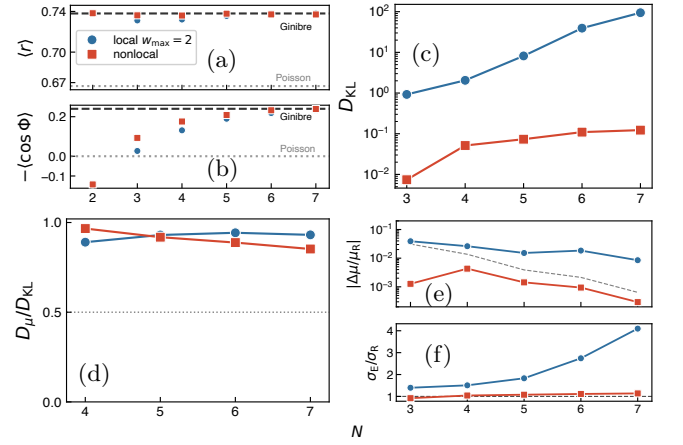

We begin with the purely dissipative models described in Sec.~\ref{sec:random-lindbladian}. We first establish their quantum-chaotic behavior using spectral statistics, focusing on the complex spacing ratio (CSR) introduced in Sec.~\ref{sec:csr}. In Fig.~\ref{fig:model-a-csr-dkl}(a)\&(b), we show the average radial and angular components of the CSR for random Lindbladians with either local or nonlocal jump operators. The radial component, $\langle r\rangle$, is already close to the GinUE value at relatively small system sizes. By contrast, the angular component, $-\langle {\rm cos}\phi\rangle$, exhibits much stronger finite-size effects and shows noticeable deviations from the GinUE prediction at small $N$. As the system size increases, however, it eventually approaches the GinUE value. These results confirm that the spectral statistics of both random Lindbladian models are consistent with quantum-chaotic behavior.

\subsection{Eigenoperator entanglement distribution}

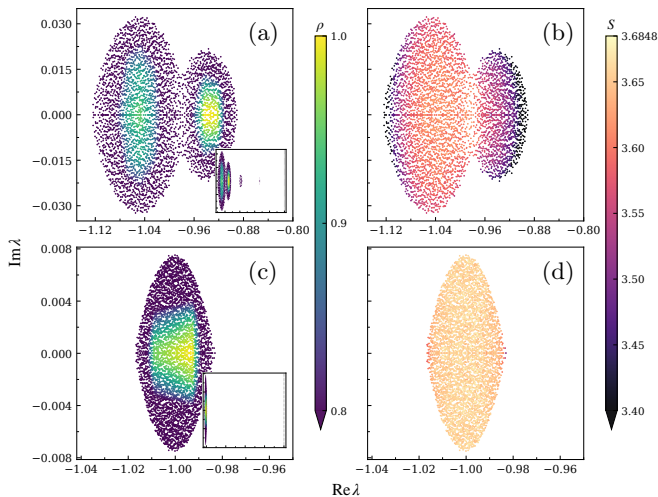
\begin{figure}[t!]
    \centering
        \begin{overpic}[width=\columnwidth]{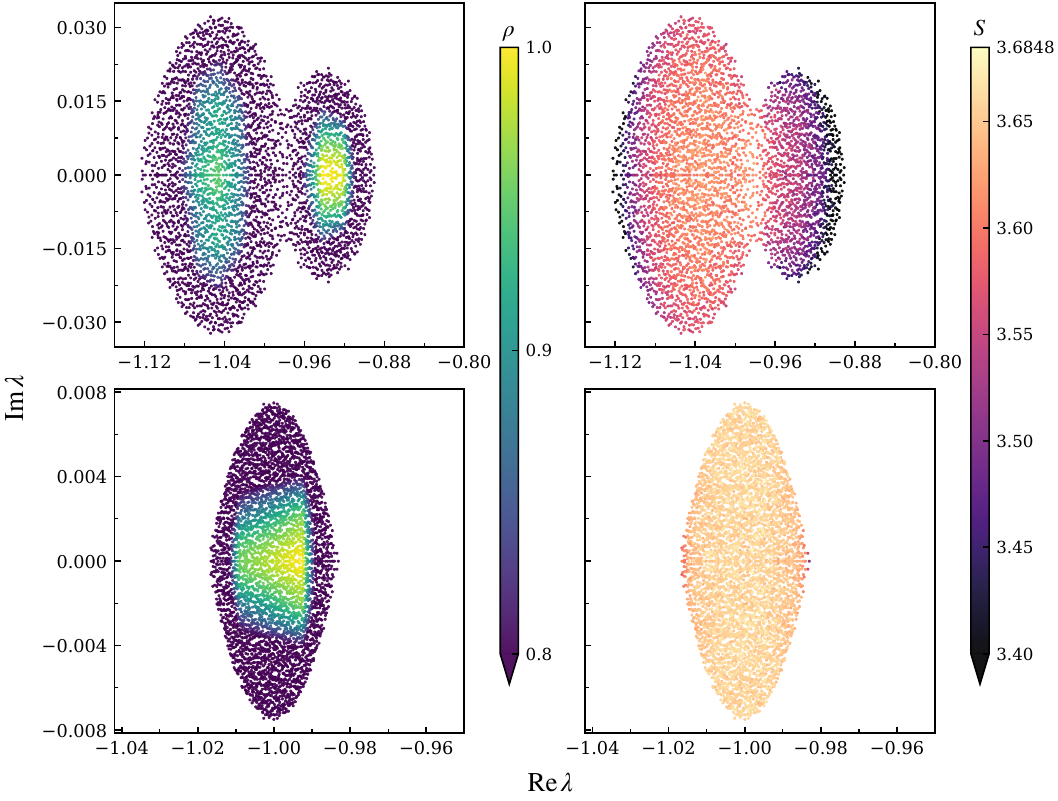}
            \put(37,70.5){\small (a)}
            \put(32,44){\includegraphics[width=0.11\columnwidth]{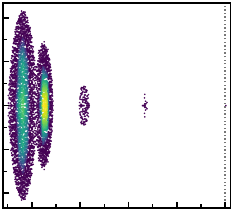}}
            \put(82,70.5){\small (b)}
            \put(37,34){\small (c)}
            \put(82,34){\small (d)}
            \put(30,8){\includegraphics[width=0.13\columnwidth]{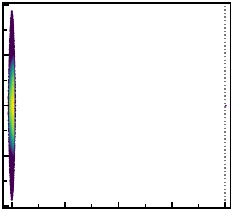}}
        \end{overpic}
    \caption{Complex spectrum structure and eigenoperator entanglement for a representative realization of the local (top row) and nonlocal (bottom row) random Liouvillians at $N=6$. The main panels enlarge the dominant spectral region, while the insets in (a) and (c) show the full spectra. (a)\&(c) Eigenvalues colored by the normalized spectral density $\rho$. (b)\&(d) The same eigenspectra colored by the half-chain operator entanglement entropy $S_A$ of the corresponding normalized right eigenoperators. Notice that locality leads to a clustered structure of the eigenspectrum, and regions with the largest density of states and highest entanglement do not coincide.}
    \label{fig:model-a-spectrum-entropy}
\end{figure}

Having established spectral chaos in random Lindbladians, we now turn to the entanglement distributions of the Liouvillian eigenoperators. In closed quantum systems, eigenstates are commonly selected within a microcanonical window near the largest density of states~\cite{EntanglementVedika}, as they are typically also states with maximal entanglement. A generic Liouvillian spectrum instead occupies the complex plane and provides no
canonical one-dimensional energy coordinate on which to define an analogous
window. Moreover, locality produces a clustered structure in the spectrum of
the local random Lindbladian~\cite{Wang2020}, as shown in
Fig.~\ref{fig:model-a-spectrum-entropy}(a). Because eigenoperators associated with different spectral clusters have dominant support at different operator weights~\cite{Wang2020, ChirameBurnell2026}, the entanglement entropy is likewise nonuniform across the complex plane. Figs.~\ref{fig:model-a-spectrum-entropy}(a) and (b) show that the region with the largest density of states need not coincide with the region of largest entanglement entropy. At this point, the choice of spectral region from which to collect entanglement statistics appears rather arbitrary and model-dependent. Moreover, for such nonuniform spectra, the mean and variance of the entanglement entropy can depend sensitively on the chosen spectral window and are therefore not robust. To overcome these difficulties, we introduce an iterative procedure inspired by the well-known $\sigma$-clipping method in statistics, which allows us to systematically extract the entanglement distribution in a model-independent manner without preselecting a spectral window.

\subsubsection{$\sigma$-clipping scheme}
\label{sec:clipping}

The standard $\sigma$-clipping procedure is a simple iterative outlier-rejection scheme. At each iteration, data points lying more than a prescribed number of standard deviations from the current mean are removed, after which the mean and standard deviation are recomputed. The procedure is repeated until the selected set converges, providing a simple way to isolate the dominant distribution while suppressing outliers. Since our goal is to extract the eigenoperator-entanglement distribution in the high-entanglement regime, we introduce an analogous iterative scheme that progressively removes low-entanglement outliers.

For realization $s$, let $S_i^{(s)}$ denote the entanglement entropy of the $i$-th eigenoperator and define
\begin{equation}
\begin{aligned}
S_{\max}^{(s)}&=\max_i S_i^{(s)},&
x_i^{(s)}&=S_{\max}^{(s)}-S_i^{(s)}\geq 0.
\end{aligned}
\end{equation}

Thus, within each realization, we take the maximum entanglement entropy as the origin and measure the distance of every other eigenoperator from this maximum. We then compute the root-mean-square distance
\begin{equation}
\Delta_0^{(s)}=\sqrt{\mathbb{E}[(x^{(s)})^2]}.
\end{equation}
At the $n$-th iteration, we retain only those eigenoperators satisfying
\begin{equation}
x_i^{(s)}\leq k\Delta_{n-1}^{(s)},
\qquad k>1,
\end{equation}
and update the root-mean-square distance according to
\begin{equation}
\left[\Delta_n^{(s)}\right]^2
=
\frac{
\displaystyle
\sum_{i:\,x_i^{(s)}\leq k\Delta_{n-1}^{(s)}}
\left(x_i^{(s)}\right)^2
}{
\displaystyle
\#\left\{i:x_i^{(s)}\leq k\Delta_{n-1}^{(s)}\right\}
}.
\end{equation}
The iteration terminates once the retained set no longer changes, at which point $\Delta_n^{(s)}$ converges to a fixed-point value $\Delta_*^{(s)}$.
For the rest of the work, we use $k=\sqrt{3}$, and we show in Appendix~\ref{sec:appendix-a} that the results are robust against other choices of $k$.
For each system size, the
retained samples are then pooled over realizations to determine
$\mu_{\mathrm{E}}$ and $\sigma_{\mathrm{E}}^2$, which enter the
Gaussian-approximation KL divergence.

\subsubsection{Entanglement distribution}

Using the iterative scheme described above, we determine the mean and variance of the high-entanglement distribution for both nonlocal and local random Lindbladians, and compare them with those of the reference Page distribution by evaluating $D_{\rm KL}$ in Eq.~(\ref{DKL-gaussian}). The results are shown in Fig.~\ref{fig:model-a-csr-dkl}(c). For nonlocal Lindbladians, $D_{\rm KL}$ remains of order $10^{-1}$ or smaller, with only a weak upward trend as the system size increases, and appears to approach saturation at the largest accessible sizes. In contrast, for local Lindbladians, $D_{\rm KL}$ increases by approximately two orders of magnitude with system size and shows no sign of saturation. This demonstrates that the near-maximally entangled eigenoperators of local Lindbladians deviate strongly from Haar-random vectors, despite exhibiting complex level-spacing statistics consistent with the Ginibre ensemble. In Appendix~\ref{sec:appendix-left-eigenoperator}, we show similar results for the left eigenoperators in both cases.

\begin{figure}[!t]
    \centering
    \subfigure{
        \begin{overpic}[width=\columnwidth]{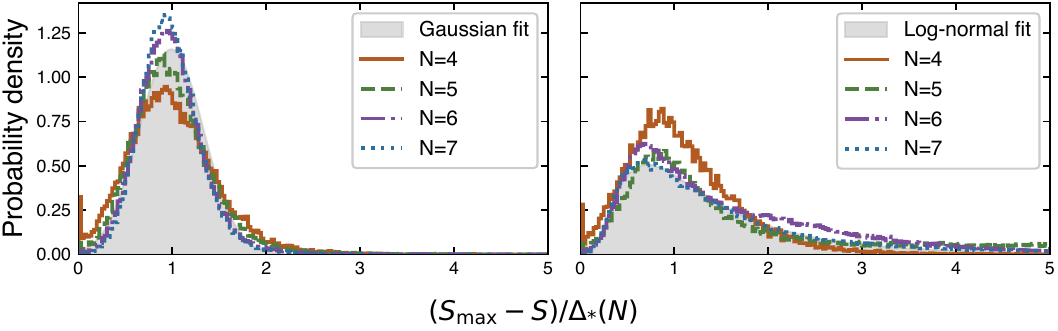}
            \put(7.8,27){\small (a)}
            \put(56,27){\small (b)}
        \end{overpic}
    }
    \subfigure{
        \begin{overpic}[width=\columnwidth]{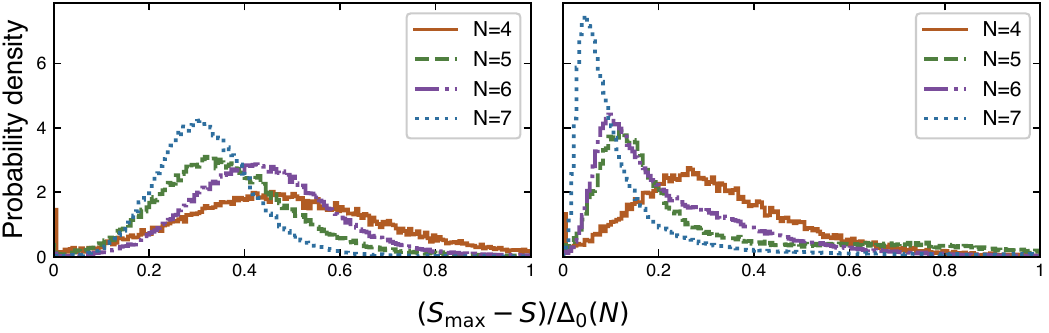}
            \put(7,27){\small (c)}
            \put(59,27){\small (d)}
        \end{overpic}
    }
    \caption{Finite-size collapse of the $\sigma$-clipped distributions in terms of the normalized variable of Eq.~(\ref{eq:normalize}), for system sizes $N=4,\ldots,7$. Panels (a) and (b) show the nonlocal and local ensembles, respectively. The gray shading shows the Gaussian fit in (a) and the log-normal fit in (b), both fitted to the $N=7$ data; the colored lines show the numerical results for $N=4,\ldots,7$. The distribution converges in both cases as the system size increases. 
    As a comparison, in (c)\&(d) we plot the raw entanglement distributions, normalized by the corresponding raw root-mean-square distance $\Delta_0(N)$. The raw distribution instead shows no clear sign of convergence with system size.}
    \label{fig:distribution-EE-core}
\end{figure}

To identify the dominant contribution to the large $D_{\rm KL}$, in Fig.~\ref{fig:model-a-csr-dkl}(d) we plot the fraction of $D_{\rm KL}$ arising from the mean contribution, $D_{\mu}$. We find that this fraction remains of order unity over the simulated system sizes, indicating that the deviation of the mean entanglement $\mu_E$ from the Page value $\mu_R$ is the primary source of the large $D_{\rm KL}$. For the nonlocal model, fraction $D_\mu/D_{\mathrm{KL}}$ decreases slowly with system size, whereas for local Lindbladians it increases weakly.
On the other hand, in $D_{\mu}$, the deviation of the mean, $\Delta\mu=|\mu_E-\mu_R|$, is normalized by the width of the Page distribution, $\sigma_R\sim 4^{-N/2}$, which is itself exponentially small in system size. It is therefore possible for $\mu_E$ to approach $\mu_R$ with increasing system size while the residual deviation is amplified by the exponentially small scale $\sigma_R$. To determine whether this is the case, we plot $|\Delta\mu/\mu_R|$ in Fig.~\ref{fig:model-a-csr-dkl}(e). For the nonlocal model, $\Delta\mu$ indeed decreases exponentially with system size, indicating that the mean entropy becomes exponentially close to the Page value. However, its decay rate is approximately the same as that of $\sigma_R$ [see the dashed line in Fig.~\ref{fig:model-a-csr-dkl}(e)], resulting in a sizable but slowly decreasing fraction $D_{\mu}/D_{\mathrm{KL}}$. In contrast, for local Lindbladians, $\Delta\mu/\mu_R$ nearly saturates to a constant, suggesting that the mean entropy does not converge to the Page value in the thermodynamic limit. We further examine the ratio of standard deviations, $\sigma_E/\sigma_R$, and find that it remains close to unity for the nonlocal model but grows rapidly with $N$ for the local model. These results show that locality of the jump operators drastically modifies the eigenoperator-entanglement distribution, producing a systematic shift in the mean together with a substantially broader distribution than the Page prediction.

Our iterative clipping scheme in fact yields the full high-entanglement distribution upon convergence, allowing us to examine the distribution itself beyond its first two moments. From the fixed-point root-mean-square distances $\Delta_*^{(s)}$, we further average over independent realizations,
\begin{equation}
\Delta_*(N)=\frac{1}{N_{\mathrm{r}}}
\sum_{s=1}^{N_{\mathrm{r}}}\Delta_*^{(s)},
\end{equation}
where $N_{\mathrm{r}}$ is the number of realizations. We then combine the retained data from all realizations, normalize them by $\Delta_*(N)$, and express the distribution in terms of the dimensionless variable
\begin{equation}
\widetilde{x}_i^{(s)}
=
\frac{S_{\max}^{(s)}-S_i^{(s)}}{\Delta_*(N)}.
\label{eq:normalize}
\end{equation}
In Fig.~\ref{fig:distribution-EE-core}, we apply this procedure and plot the resulting distributions. For both nonlocal and local Lindbladians, the distributions converge with increasing system size, indicating a well-defined high-entanglement distribution in the thermodynamic limit [Figs.~\ref{fig:distribution-EE-core}(a)\&(b)]. To highlight the importance of the iterative clipping procedure, in Figs.~\ref{fig:distribution-EE-core}(c)\&(d) we instead plot the \textit{raw} entanglement distributions, normalized by the corresponding \textit{raw} root-mean-square distance $\Delta_0(N)$. In this case, the distributions show no clear sign of convergence with increasing system size. This comparison demonstrates that the iterative scheme is essential for systematically removing model-dependent low-entanglement outliers and isolating the asymptotic distribution in the high-entanglement regime.

Remarkably, we find that the eigenoperator-entanglement distribution falls into distinct forms once locality is imposed. For nonlocal models, the asymptotic distribution is well approximated by a Gaussian, similar to the Haar-random ensemble. In contrast, for local Lindbladians the distribution is extremely well described by a log-normal form and is strongly non-Gaussian. This is consistent with the increasing $D_{\rm KL}$ and the substantially broader variance observed for the local models. Thus, unlike in Hamiltonian systems, imposing locality on the jump operators qualitatively changes the high-entanglement part of the eigenoperator distribution. The convergence with $N$ in Fig.~\ref{fig:distribution-EE-core}(b) further suggests that the log-normal form is asymptotically robust for local Lindbladians and distinct from the Gaussian behavior of random-matrix ensembles and nonlocal Lindbladians.

\section{Mixed-field Ising chain with local dissipators}\label{sec:Physical-Models}

\begin{figure}[t!]
    \centering
    \begin{overpic}[width=0.99\columnwidth]{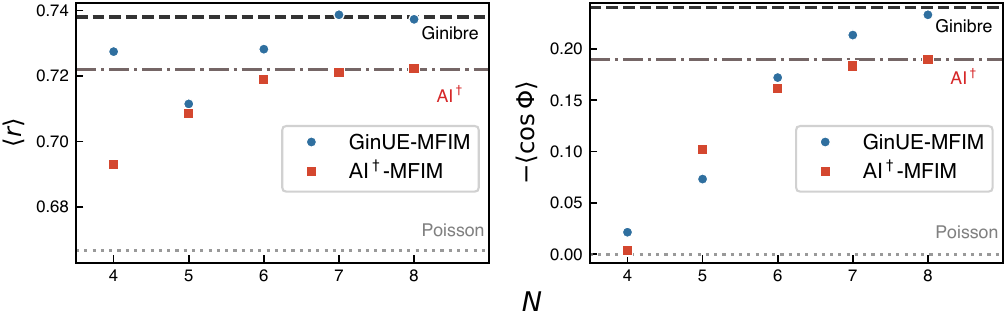}
        \put(9,11){\small (a)}
        \put(60,11){\small (b)}
    \end{overpic}
    \caption{Complex-spacing-ratio statistics for the mixed field Ising chain with dissipators~(\ref{eq:ginue})
    (blue circles) and~(\ref{eq:AI}) (red squares), respectively. Horizontal lines mark the
    GinUE, class-$\mathrm{AI}^{\dagger}$ Gaussian, and Poisson reference values. With increasing
    system size, they saturate to RMT predictions for the GinUE and class-$\mathrm{AI}^{\dagger}$ ensembles, respectively. }
    \label{fig:model-b-csr}
\end{figure}

In the previous section, we found that the eigenoperator-entanglement distribution is drastically altered once locality is imposed on the jump operators. Its convergence with increasing system size suggests that the log-normal distribution observed for local Lindbladians may represent a distinct universality class of local Liouvillians. To further test this possibility in more physical settings, we now turn to the mixed-field Ising chain with local dissipators introduced in Sec.~\ref{sec:ising}, which incorporates coherent Hamiltonian dynamics and allows us to examine models belonging to distinct symmetry classes.

\subsection{Spectral statistics}

 In parallel with the random-Liouvillian models, we first establish that, for our choice of parameters, the dissipative mixed-field Ising model exhibits spectral chaos consistent with random-matrix predictions for the corresponding symmetry class. In Fig.~\ref{fig:model-b-csr}, we show the CSR statistics for the dissipators in Eqs.~(\ref{eq:ginue}) and~(\ref{eq:AI}), respectively. Compared with the random-Lindbladian models, we observe more pronounced finite-size effects in both the radial and angular moments. Nevertheless, with increasing system size, both quantities converge toward the RMT predictions for the GinUE and class-${\rm AI}^\dagger$ ensembles, respectively.

\begin{figure}[!t]
    \centering
    \begin{overpic}[width=0.98\columnwidth]{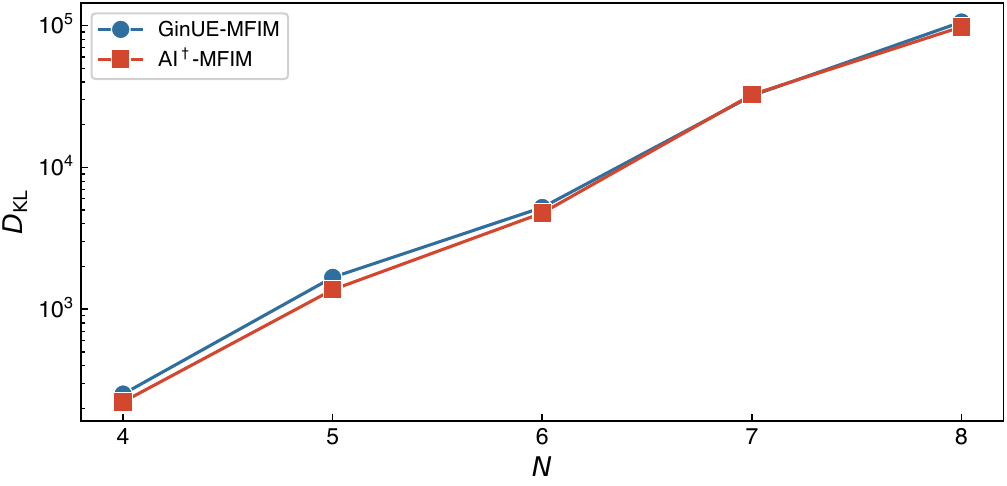}
        \put(93,10){\small(a)}
    \end{overpic}
    \begin{overpic}[width=0.98\columnwidth]{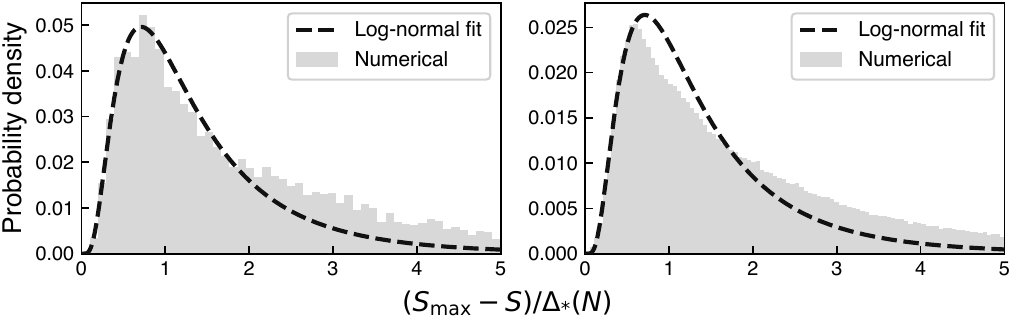}
        \put(43,8.7){\small(b)}
        \put(93,8.7){\small(c)}
    \end{overpic}

    \caption{Eigenoperator-entanglement distribution for the dissipative mixed-field Ising chain. (a) Gaussian-approximation $D_{\mathrm{KL}}$ relative to the finite-size Page reference as a function of $N$. (b),(c) Normalized $\sigma$-clipped entanglement-entropy distributions for dissipators~(\ref{eq:ginue}) and~(\ref{eq:AI}), respectively, at $N=7$. Gray histograms show the numerically obtained distributions. The dashed log-normal curves are taken directly from the fit in Fig.~\ref{fig:distribution-EE-core}, with the same parameters and no additional fitting.}
    \label{fig:model-b-eigenoperator-diagnostics}
\end{figure}

\subsection{Eigenoperator entanglement statistics}

We now turn to the eigenoperator-entanglement distributions. As in the local random-Lindbladian model, the eigenvalues of the dissipative mixed-field Ising model again exhibit a clustered structure in the complex plane. We therefore apply the same $\sigma$-clipping scheme introduced in Sec.~\ref{sec:clipping}. In Fig.~\ref{fig:model-b-eigenoperator-diagnostics}(a), we plot the Gaussian-approximation $D_{\rm KL}$ for both models and find that it increases rapidly with system size in both cases, closely paralleling the behavior of the local random-Lindbladian model. In Appendix~\ref{sec:appendix-model-b}, we further show that the mean contribution $D_\mu$ again accounts for most of the deviation.

In Figs.~\ref{fig:model-b-eigenoperator-diagnostics}(b)\&(c), we plot the full entanglement distributions obtained after convergence of the $\sigma$-clipping procedure. Remarkably, both distributions collapse onto the \textit{same} log-normal distribution obtained from the local random-Lindbladian model, using exactly the same parameters and without any additional fitting. This agreement persists even though the model in Eq.~(\ref{eq:AI}) belongs to a distinct non-Hermitian symmetry class. These results provide strong support for our conjecture that the log-normal distribution is robust to microscopic details and symmetry class, and may represent a universal feature of physical Lindbladians with locality.

\section{Summary and Outlook}
\label{sec:conclsn}

In this work, we showed that spectral chaos in open quantum systems does not imply Haar-typical Liouvillian eigenoperators. Although local and nonlocal Lindbladians can exhibit the same random-matrix spectral statistics, their eigenoperator-entanglement distributions behave very differently. Nonlocal Lindbladians remain close to Haar-random behavior, whereas locality produces strong and growing deviations with system size. Using an iterative $\sigma$-clipping scheme to isolate the high-entanglement regime, we further found a Gaussian distribution for nonlocal Lindbladians but a robust log-normal distribution for local ones. Remarkably, the same log-normal distribution, with no additional fitting, also describes dissipative mixed-field Ising chains in distinct non-Hermitian symmetry classes, pointing to a universal eigenoperator-entanglement structure associated with locality.

An important open question is the microscopic origin of this log-normal distribution. It would be particularly interesting to understand whether it can be derived from locality, operator growth, or other structural constraints on Liouvillian eigenoperators. It also remains to be determined how broadly this behavior extends to other local open quantum systems, including models with conservation laws, higher spatial dimensions, and different forms of dissipation. More generally, our results suggest that eigenoperator statistics can reveal universal structure beyond what is visible from eigenvalue correlations alone, providing a complementary perspective on quantum chaos in open many-body systems.

\begin{acknowledgments}
This work is supported by Grant No. 12375027 from the National Natural Science Foundation of China.
Numerical simulations were performed on the High-performance
Computing Platform of Peking University.
\end{acknowledgments}

\appendix
\renewcommand{\theequation}{S\arabic{equation}}
\setcounter{equation}{0}
\renewcommand{\thefigure}{S\arabic{figure}}
\setcounter{figure}{0}
\renewcommand{\theHequation}{supp.\arabic{section}.\arabic{equation}}
\renewcommand{\theHfigure}{S\arabic{figure}}
\providecommand{\theHsubfigure}{}
\renewcommand{\theHsubfigure}{S\arabic{figure}.\arabic{subfigure}}

\section{Robustness of the $\sigma$-clipping prescription}
\label{sec:appendix-a}

\begin{figure}[t!]
    \centering
    \begin{overpic}[width=\columnwidth]{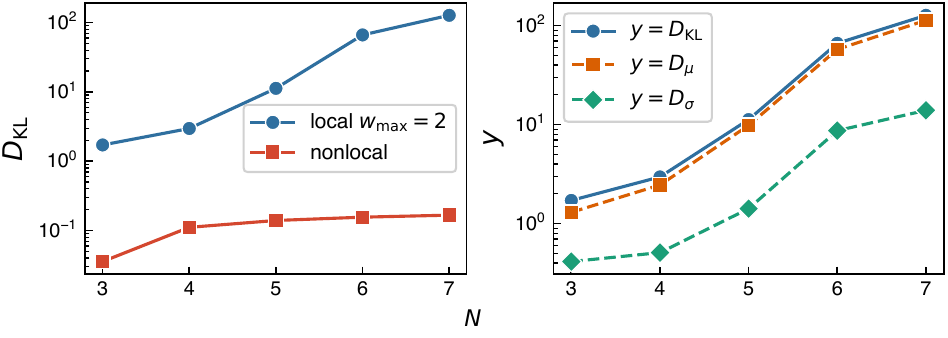}
        \put(11.5,12){\small (a)}
        \put(61,17){\small (b)}
    \end{overpic}
    \begin{overpic}[width=0.98\columnwidth]{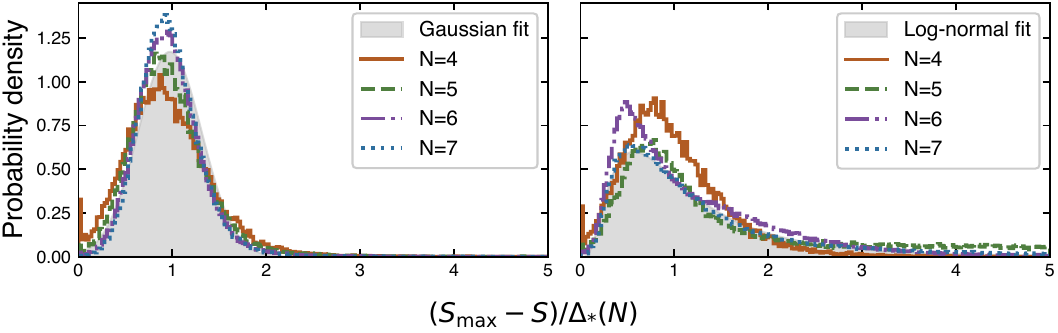}
        \put(44,9.2){\small(c)}
        \put(93,9.2){\small(d)}
    \end{overpic}
    \caption{(a) $D_{\rm KL}$ for the random Lindbladian models obtained using a different choice $k=2$ in the $\sigma$-clipping procedure. (b) Decomposition of the local
    result into the mean contribution $D_{\mu}$ and the variance contribution
    $D_{\sigma}$. The results are consistent with the $k=\sqrt{3}$ analysis in
    Figs.~\ref{fig:model-a-csr-dkl}(c) and \ref{fig:model-a-csr-dkl}(d). (c), (d) The $\sigma$-clipped entanglement distribution for the nonlocal and local ensembles, respectively, the counterparts of Fig.~\ref{fig:distribution-EE-core}(a)\&(b). }
    \label{fig:DKL-k=2}
\end{figure}
In this section, we show that our $\sigma$-clipping scheme is robust against different choices of $k$, provided that it is sufficiently far from unity. In the main text, we choose $k=\sqrt{3}$. In Fig.~\ref{fig:DKL-k=2}(a)\&(b), we show numerical results of $D_{\rm KL}$ computed for the random Lindbladians with a different choice $k=2$, and in (c)\&(d) the corresponding $\sigma$-clipping distributions. We find that the results are qualitatively the same as in  Fig.~\ref{fig:model-a-csr-dkl} and Fig.~\ref{fig:distribution-EE-core} in the main text.

\section{Left-eigenoperator entanglement statistics for random Lindbladians}
\label{sec:appendix-left-eigenoperator}

\begin{figure}[t!]
    \centering
    \subfigure{

        \begin{overpic}[width=0.99\columnwidth]{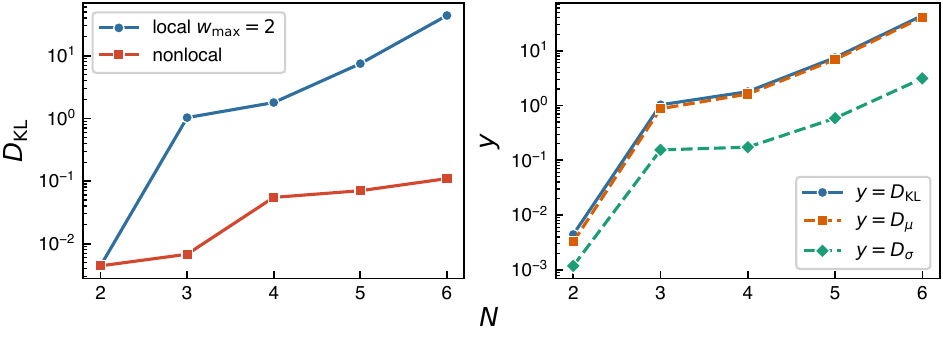}
            \put(44,8){\small (a)}
            \put(80,8){\small (b)}
        \end{overpic}
    }
    \subfigure{
        \begin{overpic}[width=0.99\columnwidth]{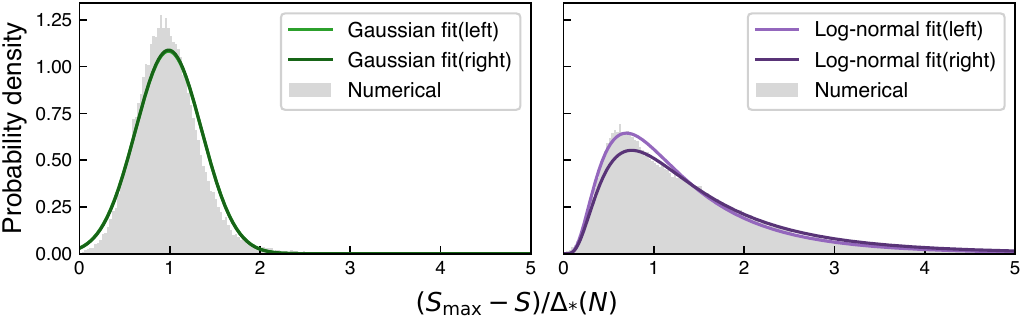}
            \put(45,8){\small (c)}
            \put(93,8){\small (d)}
        \end{overpic}
    }
    \caption{Eigenoperator entanglement statistics obtained using left eigenoperators of random Lindbladian models. (a) Gaussian-approximation $D_{\mathrm{KL}}$ of the local (blue circles) and nonlocal (red squares) ensembles. (b) Decomposition of $D_{\mathrm{KL}}$ for the local ensemble into the mean $D_\mu$ and variance $D_{\sigma}$ contributions. (c) and (d) show the normalized $\sigma$-clipped entanglement distributions of the nonlocal and local ensembles, respectively, at $N=6$, together with the corresponding Gaussian and log-normal fits. For comparison, the darker curves are direct fits to right-eigenoperator data, while the lighter curves are direct fits to left-eigenoperator data shown.}
    \label{fig:model-a-left-dkl}
\end{figure}

In this section, we show that the left-eigenoperators exhibit similar statistical properties as the right-eigenoperators studied in the main text. We restrict ourselves to the random Lindbladian models. In Fig.~\ref{fig:model-a-left-dkl}, we show the numerical results for both the Gaussian approximation $D_{\rm KL}$ and the $\sigma$-clipped entanglement distributions. Again, we find that the $D_{\rm KL}$ obtained using left eigenoperators follows qualitatively the same trend as the right eigenoperators, with main contributions also coming from deviations in the mean value. Moreover, the $\sigma$-clipped distribution similarly follows a Gaussian and log-normal distribution for the nonlocal and local ensembles, respectively. For nonlocal ensemble, the fitted curves for left and right eigenoperators are nearly identical, whereas for the local ensemble they differ visibly in their parameters, but retain the same log-normal shape.

\section{KL-divergence decomposition for the dissipative mixed-field Ising chain}
\label{sec:appendix-model-b}

\begin{figure}[t!]
    \centering
        \begin{overpic}[width=\columnwidth]{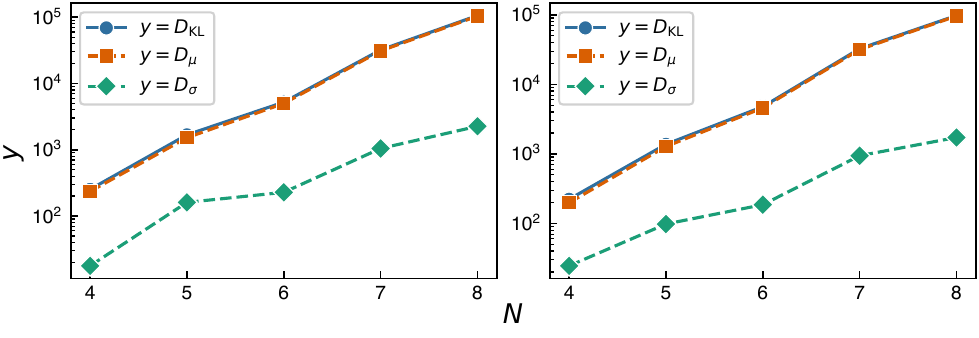}
        \put(42.5,9){\small (a)}
        \put(93,9){\small (b)}
        \end{overpic}
    \caption{Decomposition of the KL divergence for the dissipative mixed-field Ising model into the mean $D_{\mu}$ and variance $D_{\sigma}$ contributions.
    (a) mixed-field Ising model with the four-channel
    dissipators in Eq.~(\ref{eq:ginue}), and (b) local
    $Z$-dephasing channels in Eq.~(\ref{eq:AI}). 
    In both cases, $D_{\mu}$ provides the dominant contribution, while $D_{\sigma}$ remains subleading.}
    \label{fig:model-b-dkl-decomposition}
\end{figure}

In Fig.~\ref{fig:model-b-dkl-decomposition}, we decompose the Gaussian-approximation $D_{\rm KL}$ for the dissipative mixed-field Ising model into the mean $D_{\mu}$ and variance $D_{\sigma}$ contributions. Again, we find that $D_{\rm KL}$ is dominated by the mean contribution, suggesting that the main source of discrepancies from the Page distribution comes from the deviation of the mean entropy.

\section{Additional numerical results on entanglement distribution}
\label{sec:appendix-additional-distributions}

\begin{figure*}[t]
    \centering
    \includegraphics[width=\textwidth]{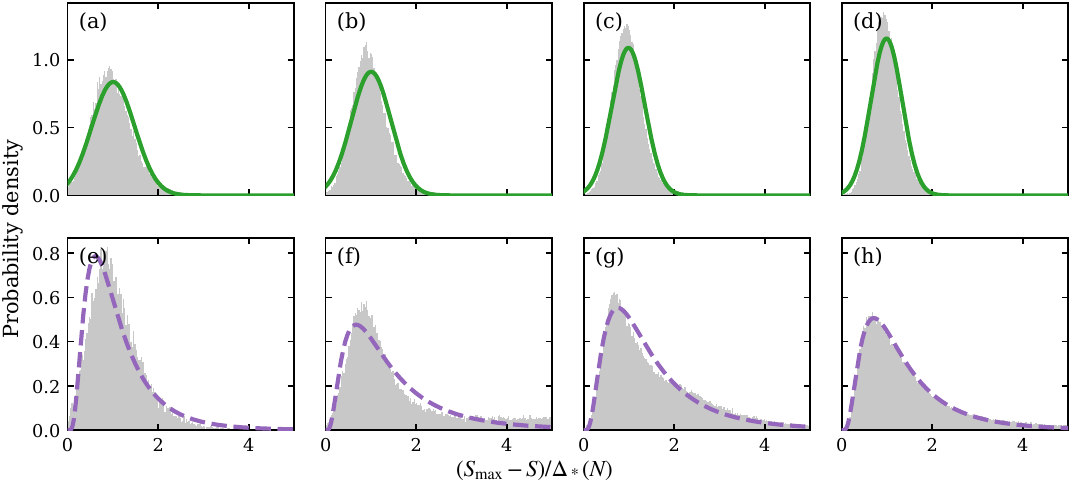}
    \caption{Finite-size fits of the normalized $\sigma$-clipped distributions of random Lindbladian models, with $x=(S_{\max}-S)/\Delta_*$. The upper row, panels (a)--(d), shows the nonlocal ensemble fitted by Gaussians; the lower row, panels (e)--(h), shows the local ensemble fitted by the log-normal. Within each row, panels run from $N=4$ to $N=7$ from left to right. Gray histograms show the numerically obtained distributions, while the solid green and dashed purple curves show the Gaussian and log-normal maximum-likelihood fits, respectively.}
    \label{fig:finite-size-core-fits}
\end{figure*}

We provide a numerical fit of the $\sigma$-clipped distributions for random Lindbladians models for all system sizes ranging from $N=4$ to $N=7$, as shown in Fig.~\ref{fig:finite-size-core-fits}. Detailed fitting parameters are listed in Table~\ref{tab:finite-size-core-fit-parameters}.

\makeatletter
\setlength{\@dblfptop}{0pt}
\setlength{\@fptop}{0pt}
\makeatother

\begin{table}[h!]
    \caption{Maximum-likelihood fit parameters for Fig.~\ref{fig:finite-size-core-fits}. The nonlocal distributions are fit by Gaussians with parameters $(\mu_{\mathrm{G}},\sigma_{\mathrm{G}})$, and the local distributions by log-normal with parameters $(\mu_{\log},\sigma_{\log})$. All fits use $0<x\leq5$.}
    \centering
    \begin{ruledtabular}
    \begin{tabular}{c c c c c}
        $N$ & \multicolumn{2}{c}{nonlocal Gaussian} & \multicolumn{2}{c}{local log-normal} \\
          & $\mu_{\mathrm{G}}$ & $\sigma_{\mathrm{G}}$ & $\mu_{\log}$ & $\sigma_{\log}$ \\
        4 & 1.0073 & 0.4752 & -0.0768 & 0.6368 \\
        5 & 1.0046 & 0.4371 & 0.1685 & 0.7501 \\
        6 & 0.9884 & 0.3664 & 0.2094 & 0.7022 \\
        7 & 0.9889 & 0.3447 & 0.1313 & 0.6878 \\
    \end{tabular}
    \end{ruledtabular}
    \label{tab:finite-size-core-fit-parameters}
\end{table}

\clearpage
\bibliography{ref}
\end{document}